\documentclass[referee]{raa}            

\usepackage{caption,subcaption}
\usepackage{graphicx,times}          
\usepackage{natbib}
\usepackage{overpic}
\graphicspath{{./}{figures/}}
\usepackage{amssymb,amsmath}
\bibpunct{(}{)}{;}{a}{}{,}
\usepackage{hyperref}
\hypersetup{colorlinks,				
	linkcolor=blue,
	citecolor=blue}

\begin{document}

   \title{Spatial Modulation Search Applied to the Search and Confirmation of Highly Scintillated Pulsars at FAST with A Pulsar Discovered in M3
}

   \volnopage{Vol.0 (20xx) No.0, 000--000}      
   \setcounter{page}{1}          

   \author{Lei Qian
      \inst{1,2,3},
      Zhichen Pan
      \inst{1,2,3},
   }

   \institute{National Astronomical Observatories, Chinese Academy of Sciences,
             Beijing 100012, China; {\it pancz@nao.cas.cn}\\
        \and
    CAS Key Laboratory of FAST, National Astronomical Observatories, Chinese Academy of Sciences, Beijing 100101, China
        \and
     University of Chinese Academy of Sciences, Beijing 100049, China \\
\vs\no
   {\small Received~~20xx month day; accepted~~20xx~~month day}}

\abstract{
We present a pulsar candidate identification and confirmation procedure
based on a position-switch mode during the pulsar search observations. This method enables the simultaneous search and confirmation of a pulsar in a single observation, by utilizing the different spatial features of a pulsar signal and a radio frequency interference (RFI).
Based on this method, we performed test pulsar search observations in globular clusters M3, M15, and M92. We discovered and confirmed a new pulsar, M3F, and detected the known pulsars M3B, M15 A to G (except C), and M92A.
}

   \authorrunning{Lei Qian, Zhichen Pan}            
   \titlerunning{Spatial Modulation Search}  

   \maketitle

\section{Introduction}           
\label{sect:intro}

In the pulsar search,
one key step is to pick pulsar signals out of an ocean of radio frequency interferences (RFIs).
Beside the dispersion phenomenon, there are at least two main differences between a pulsar signal and an RFI.
The first difference is temporal.
Usually, a pulsar without very severe scintillation is more persistent than an RFI,
in the sense that this pulsar will appear in different observations while a particular RFI usually appears in only  one observation.
The second difference is spatial. A pulsar is a point source, whose signal goes into the main beam while an RFI is usually picked up by the side-lobes.
So the RFI does not vary much when the telescope moves away from a pulsar.

Currently, the confirmation of pulsar candidates was normally done by re-observing in the same direction, i.e.
making use of the temporal differences between a pulsar signal and an RFI.
However, it would be difficult for this strategy to be applied to the highly scintillated candidates
or candidates with possibly significant acceleration.
Since it would require several observations to re-detect such pulsars,
i.e. it takes a longer time to verify or falsify such candidates.
As an example, the Globular Cluster (GC) pulsar M3C and NGC6749B were detected only once in the former tens of observations \citep{2005AAS...207.3205R}, and still not confirmed.
The extreme example is the GC pulsar 47 Tuc aa,
detected with a statistic-based search method \citep{2016MNRAS.459L..26P}.
It has a very low detection probability \citep[4.2\% or a detection rate of 1.38 detections per year,][]{2018MNRAS.476.4794F}.
The searching and confirmation of these pulsar candidates need much observation time.

The identification of pulsar candidates are normally based on the pulse profile, time-domain features,
frequency domain features, and the dispersion measure (DM) features.
Although artificial intelligence helps a lot in the identification of pulsar candidates \citep{2014ApJ...781..117Z},
all the candidates still need to be re-observed for confirmation or falsification.
In this paper, inspired by the position-switch observing mode,
we introduced the spatial modulation search,
making use of the spatial differences between a pulsar signal and an RFI.
This method was tested with
the Five-hundred-meter Aperture Spherical radio Telescope \citep[FAST, ][]{2006ScChG..49..129N,2011IJMPD..20..989N,2019SCPMA..6259502J,2020Innov...100053Q}.
As a result, a GC pulsar, M3F was discovered and confirmed.
The data and the search method were described in section 2.
In section 3 we presented the results and made some discussion.
The conclusions are presented in section 4.

\section{Data and Methods}

Usually, a pulsar was considered confirmed if it was re-detected at least once.
Inspired by the observing mode widely used in spectral observations,
we suggest that the searching and confirmation of pulsars (distinguishing them from RFIs) can be combined into
one observation by using their spatial characteristics.
To be specific, the position-switch mode can be used.
A signal from a celestial point source, e.g. a pulsar, should disappear when we observe the OFF positions (point the telescope away from the source) and may appear again when we move the telescope back to the source (the ON position).
We kept the data recording during the whole observing process.
Thus, the pattern of appearing and disappearing of a candidate's signal can be used to distinguish a pulsar from an RFI, i.e. to confirm or exclude a pulsar candidate.

The test observations with this strategy were performed with FAST in January and February 2021,
by looking at the globular clusters M3, M15, and M92. Among these three globular clusters, M3 and M15 are chosen because of the scintillating pulsars in them, while M92 are chosen because the pulsar M92A is eclipsing for about one-third of the orbital period \citep{2020ApJ...892L...6P}, being a good example to test if the eclipsing affects this pulsar search strategy. The globular clusters M15 and M92 were observed once each.
Since the pulsars in M3 have detection probabilities lower than 50\% \citep{2007ApJ...670..363H},
we observed M3 twice in order to have more pulsar detections.
All the observations lasted for two hours, except a 5-hour observation of M3.
To optimize these test observations,
we observed the OFF position ($5'$ east to the cluster) 6 times during each 2-hour observation.
The details of the observation are listed in Table \ref{obs}.

The 19-beam receiver was used in these observations.
It covers a frequency range of 1.05 to 1.45~GHz.
The data was channelized into 4096 channels (with a channel width of 0.122~MHz).
The system temperature is $\rm \sim$24~K \citep{2020RAA....20...64J},
and the beam size is $\rm \sim 3'$ at 1.4~GHz.
The data were sampled in two polarizations with 8-bit precision every 49.152~$\mu$s.

The data were processed with PRESTO \citep{2001PhDT.......123R}.
The routines {\tt rfifind}, {\tt prepsubband}, {\tt realfft}, and {\tt accelsearch} were used for RFI masking, dedispersion,
fast Fourier transform, and acceleration search \citep{2002AJ....124.1788R}, respectively.
To catch the binary signals, we used a zmax value of 1200 for {\tt accelsearch} in all the 2-hour observations.
For the 5-hour M3 observation, a zmax of 600 was used,
due to the limitation of the memory and the computation power.
We checked all the search results to catch any possible known pulsar signals.
The details of the observations and the results were listed in Table \ref{obs}.

\begin{table*}[htpb]
\centering
\caption{Observation details and the pulsar detected.}
\label{obs}
\begin{tabular}{ccccc}
\hline
GC Name     &   Obs Date       &   Obs. Length   &  ON-OFF Time  &  Pulsar Detections  \\
            &   (YYYYMMDD)     &    (Hours)      &    (s)        &                     \\
\hline
M3          &   20210104       &   2             &  \textbf{2200}-300-\textbf{1980}-300-\textbf{1200}-300-\textbf{720}-300-\textbf{240}-300-\textbf{420}-300 & B, F\\
M3          &   20210112       &   5             &  300-\textbf{17700}                                                                                      & B, D, E \\
M15         &   20210202       &   2             &  \textbf{2200}-300-\textbf{1980}-300-\textbf{1200}-300-\textbf{720}-300-\textbf{240}-300-\textbf{420}-300 & A, B, D, E, F, G \\
M92         &   20210113       &   2             &  \textbf{2200}-300-\textbf{1980}-300-\textbf{1200}-300-\textbf{720}-300-\textbf{240}-300-\textbf{420}-300 &      A        \\
\hline
\end{tabular}

{\tiny 1$^{st}$ column: Name of the globular cluster;
2$^{nd}$ column: Observing date;
3$^{rd}$ column: Observing time length;
4$^{th}$ column: The observing cycle, with the ON observation times marked with bold font;
5$^{th}$ column: The pulsar detections.}
\end{table*}

\section{Results and Discussion}

We detected 9 pulsars among the 13 previously known pulsars in these three GCs. In the 4 undetected pulsars, M3A \citep[detected only once previously,][]{2007ApJ...670..363H}
was detected and confirmed by FAST recently \citep{Pan2021}, but not detected in this work; M3C was not detected in this work;
M15C is not detected due to either its orbital period is relatively too short compared with the observation time, or it is too faint; M15H is not detected because it is too faint.
Both M15C and M15H were discovered by Arecibo at 430 MHz with a 10 MHz bandwidth \citep{1993PhDT.........2A}.
According to the results of FAST M15 observations,
M15C, G, and H are the three pulsars most difficult to detect, due to the lowest detection rate.
This can be explained as follows.
In an eccentric and highly accelerated compact orbit, M15C is not easy to detect in the pulsar search although it is relatively bright.
M15G and H are faint for most of the time and can only be detected in a relatively bright state due to scintillations.
In the M15 observation of this work, M15G was luckily detected.
For M15H, even with FAST, it is only possible to detect it in observations of 3 hours or longer.
Among the detected pulsars, M3E was a binary pulsar previously discovered \citep{Pan2021}. It was confirmed in our test observations. Moreover, a new pulsar, M3F was discovered and confirmed in our test observations in this work.

\begin{figure*}
\centering
    \includegraphics[width=12cm]{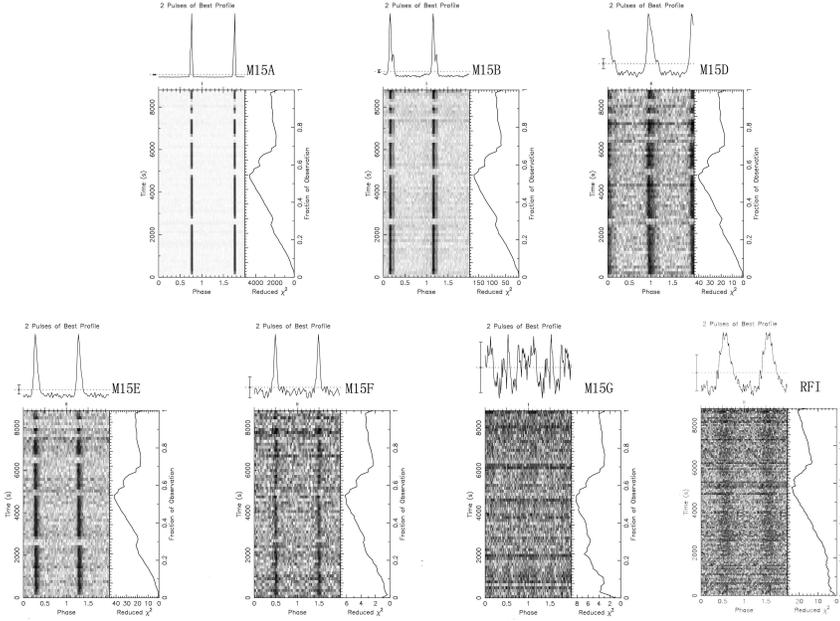}
    \caption{The examples of known pulsar detections and RFIs.
    The upper three plots are M15A, B, and D from left to right.
    The lower four plots are M15E, F, G, and an RFI.}
    \label{psr_plots_m15}
\end{figure*}

The examples of known pulsar detections and RFIs were shown in Figure \ref{psr_plots_m15}.
All the known bright pulsars have similar discontinuous vertical line patterns due to the position-switch observation.
While the RFI signal exhibits continuous vertical line patterns throughout the whole observation.
It is clear from these comparisons that the signals with such discontinuous vertical patterns can be confirmed as pulsars. We detected the pulsar M15G with luck. It is a very faint pulsar never detected by FAST before. This is an example that the spatial modulation search may not works well for extremely faint candidates when checking candidates by eyes. A possible statistical method of signal checking may be applied to find those faint signals in future works.

\begin{figure*}
\centering
    \includegraphics[width=16cm]{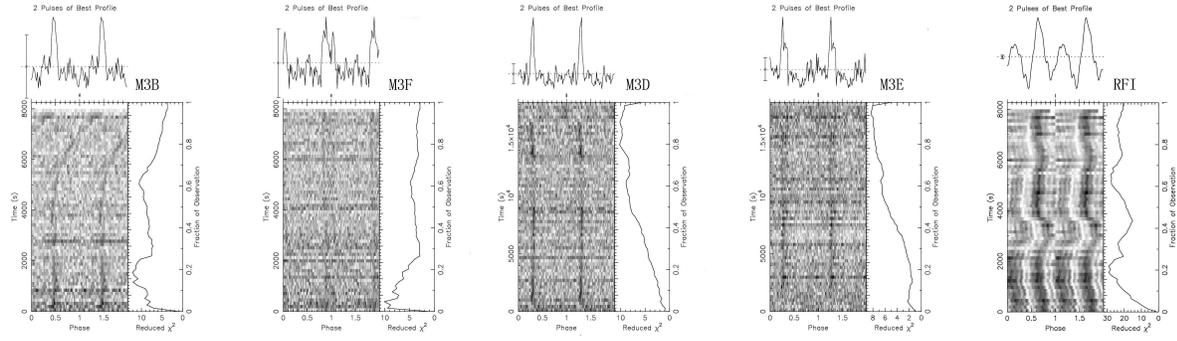}
    \caption{Pulsars in M3 and an RFI. From left to right: detection plots of M3B, F, D, E,{\bf and an RFI}.
    Though affected by the orbital movement (M3B in 20210104, the first panel) and scintillation (M3D in 20210112, the third panel), pulsars have discontinuous vertical line patterns corresponding to position switching. They are easy to be identified by eyes.}
    \label{psr_plots_m3}
\end{figure*}

Figure \ref{psr_plots_m3} shows the details of the pulsars detected in the GC M3.
In the search, we can not remove all the period variations of M3B caused by the orbital movements.
This caused the signal to noise ratio to decrease slightly,
while the time domain signal appearances were clear and similar to that of a celestial source.
The M3D signal is affected by scintillation but it is still easy to see that the signal appeared and disappeared according to the ON/OFF switches in the observation.
The pulsar M3F was detected and confirmed in the observation done on January 4$^{th}$, 2021, for the first time.
While M3E was detected previously also by FAST and confirmed in the observation on January 12$^{th}$, 2021.
Due to scintillation, the gap between the detections of either M3E or F can be as long as several months by checking the archival data. Thus, we only obtain their orbital parameters by timing them with JUMPs.
M3E is a binary pulsar in a circular orbit with an orbital period of 7.1 days and a minimum companion mass of $0.2 M_{\odot}$. M3F is a binary pulsar in a 3.0-day circular orbit with a minimum companion mass of $0.15 M_{\odot}$.

This strategy requires extra observation time for the OFF positions.
In our test, about 10\% time are used for the OFF position observation.
We estimated the searching and confirmation efficiency as follows:
  \begin{equation}
    {\rm PSR}_{\rm eff} = \frac{T_{\rm tot}}{N_{\rm psr}} = \frac{T_{\rm search} + T_{\rm confirm}}{N_{\rm psr}} = \frac{T_{\rm search} + T_{\rm obs} \times N_{\rm cand}}{N_{\rm psr}},
    \label{eq_eff1}
  \end{equation}
In which ${\rm PSR}_{\rm eff}$ is the searching and confirmation efficiency, in the unit of hours per pulsar, $T_{\rm tot}$ is the total observation time, $T_{\rm search}$ is the time used for detecting a pulsar candidate, $T_{\rm confirm}$ is the time used for candidate confirmation, $T_{\rm obs}$ is the time used to confirm or falsify a candidate, and $N_{\rm cand}$ and $N_{\rm psr}$ are numbers of candidates and pulsars, respectively. While if the spatial modulation search requests for extra time of a fraction of $x$, the searching and confirmation efficiency is:
  \begin{equation}
    {\rm PSR}_{\rm eff} = \frac{T_{\rm search} \times (1 + x)}{N_{\rm psr}},
    \label{eq_eff2}
  \end{equation}
Comparing Equation \ref{eq_eff1} and \ref{eq_eff2},
it is obvious that if the confirmation time is longer than the extra time used for the OFF position observation, the pulsar searching and confirmation efficiency are lower when using the spatial modulation search.
In the typical pulsar surveys, one pulsar discovery may need tens of hours.
On the other hand, for targeted search and confirmation of faint candidates, normally two or more observations are  needed. Thus, in general, the search efficiency could be improved by using the spatial modulation search method.

When using the position-switch observing mode, the coherence of pulsar signals in FFT analysis will be affected.
One possible way to avoid this effect is to set the OFF position observation at the beginning or the end of the observation.
Current tests on faint M3 pulsars show that adding several short OFF-position observation segments affects little in the pulsar searches.

\section{Conclusions}
\label{sect:concl}

We have performed test observations with the spatial modulation search method, with pulsars discovered and/or confirmed. With these observations, we came to the following conclusions.

1. Observing with a mode similar to the ON-OFF mode forms an artificial pattern in the phase-time diagram of a real pulsar, which is absent for an RFI.

2, We tested the spatial modulation search method with FAST, with a binary millisecond pulsar, M3E confirmed, and the other one, M3F discovered and confirmed.

3, M3E and F (J1342+2822E and J1342+2822F) are pulsars in binary systems of circular orbits, with spinning periods of 3.47 and 4.40 ms, and orbital periods of 7.1 and 3.0 days, respectively.

4, Previously known pulsars M15A, B, D, E, F, and G, and M92A were also detected, while no new pulsars were discovered in the globular clusters M15 and M92.

\

\noindent {\bf Acknowledgments} This work is supported by National SKA Program of China No. 2020SKA0120100.
Lei Qian is supported by the Youth Innovation Promotion Association of CAS (id.~2018075).
This work is supported by the Basic Science Center Project of the National Nature Science Foundation of China (NSFC) under Grant No. 11703047.
ZP is supoorted by the CAS "Light of West China" Program.
FAST is a Chinese national mega-science facility, built and operated by the National Astronomical Observatories, Chinese Academy of Sciences (NAOC).
We appreciate all the people from FAST group for their support and assistance during the observations.

\end{document}